# Inverted GaInP/GaAs Three-Terminal Heterojunction Bipolar Transistor Solar Cell

Marius H. Zehender
*Instituto de Energía Solar*
*Universidad Politécnica de Madrid*
Madrid, Spain
orcid: 0000-0002-2263-4560

Simon A. Svatek
*Instituto de Energía Solar*
*Universidad Politécnica de Madrid*
Madrid, Spain
orcid: 0000-0002-8104-1888

Myles A. Steiner
*National Renewable Energy Laboratory*
Golden, USA
orcid: 0000-0003-1643-9766

Iván García
*Instituto de Energía Solar*
*Universidad Politécnica de Madrid*
Madrid, Spain
orcid: 0000-0002-9895-2020

Pablo García Linares
*Instituto de Energía Solar*
*Universidad Politécnica de Madrid*
Madrid, Spain
orcid: 0000-0003-2369-3017

Emily L. Warren
*National Renewable Energy Laboratory*
Golden, USA
orcid: 0000-0001-8568-7881

Antonio Martí
*Instituto de Energía Solar*
*Universidad Politécnica de Madrid*
Madrid, Spain
orcid: 0000-0002-8841-709

Adele C. Tamboli
*National Renewable Energy Laboratory*
Golden, USA
orcid: 0000-0003-2839-9634

Elisa Antolín
*Instituto de Energía Solar*
*Universidad Politécnica de Madrid*
Madrid, Spain
orcid: 0000-0002-5220-2849

*Abstract*—Here we present the experimental results of an inverted three-terminal heterojunction bipolar transistor solar cell (HBTSC) made of GaInP/GaAs. The inverted growth and processing enable contacting the intermediate layer (base) from the bottom, which improves the cell performance by reducing shadow factor and series resistance at the same time. With this prototype we show that an inverted processing of a three-terminal solar cell is feasible and pave the way for the application of epitaxial lift-off, substrate reuse and mechanical stacking to the HBTSC which can eventually lead to a low-cost high-efficiency III-V-on-Si HBTSC technology.

*Keywords—double-junction, three-terminal, inverted, thin-film, III-V, solar cell, (Al)GaInP/r/GaAs(p/n)*

I. INTRODUCTION

The HBTSC [1] concept enables the fabrication of a three-terminal double-junction solar cell in which tunnel junctions and current matching are avoided. With our recent work [2], [3] we have shown that this concept has the potential to produce a high-efficiency double-junction solar cell, and now we present with the inverted structure a further milestone in the progress to design a competitive solar cell for low-cost, high-efficiency applications. This is an important step in the HBTSC technology roadmap because it enables the fabrication of large area devices and the mechanical stacking of HBTSC devices onto another type of solar cell, for example silicon cells. The HBTSC+Si tandem could make up a very high efficiency triple-junction solar cell and facilitate the interconnection within a two-terminal module [4]. Furthermore, the independence of the HBTSC from its original substrate allows the application of epitaxial lift-off and substrate reuse [5], which helps lowering substrate cost.

Fig.1 shows an upright HBTSC [2]. The core of the cell is a stack of three oppositely doped semiconductor layers, forming two *p-n* junctions in only three layers. For further reference, we call these layers, from top to bottom, emitter, base, and collector. In this way, the emitter and base form the top junction and base and collector form the bottom junction. Each of these three layers must be contacted by a terminal, making this cell a three-terminal cell. As depicted in Fig. 1, the top junction is contacted by terminals T and Z and the bottom junction by R and Z. Interestingly, the base layer does not need to be especially thick or resistive for the two junctions to deliver power independently from each other. In the case of a (Al)GaInP/GaAs HBTSC, with the correct design, a base thickness well below 1 µm is enough to prevent crosstalk between the junctions [2], [6]. This makes the HBTSC an extremely compact form of double-junction solar cell. Recently, a naming convention for three-terminal cells has been proposed [7]. Following this guideline, this cell is denominated (Al)GaInP/r/GaAs(p/n).

This work has been funded by the European Union's Horizon 2020 research and innovation program under grant agreement No. 787289. by the Spanish Science Ministry under Grant RTI2018-096937-B-C21, and by the Fundación Iberdrola within the ConCEPT II Project. M.H.Z. is grateful to Universidad Politécnica de Madrid for funding through the Predoctoral Grant Programme. I.G. and E.A. acknowledge Ramón y Cajal Fellowships (RYC-2014-15621 and RYC-2015-18539, respectively) and S.A.S acknowledges a Juan de la Cierva Fellowship (FJC2018-036517-I), funded by the Spanish Science Ministry. This work was supported by the National Renewable Energy Laboratory for the U.S. Department of Energy (DOE) under Contract No. DE-AC36-08GO28308. Funding provided by the U.S. Department of Energy, Office of Energy Efficiency and Renewable Energy, Solar Energy Technologies Office (SETO). The views expressed in the article do not necessarily represent the views of the DOE or the U.S. Government.

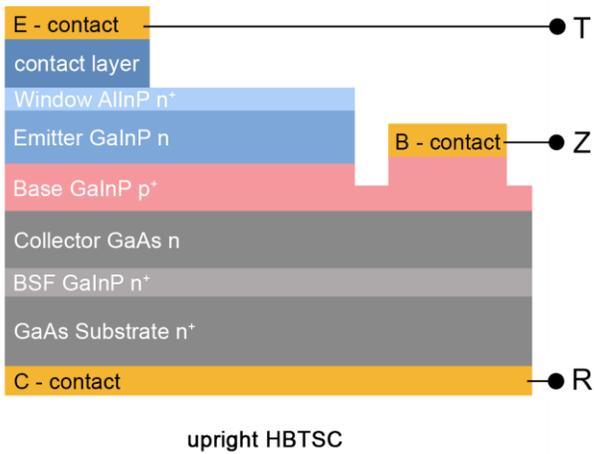

Fig. 1. Layer structure of an upright GaInP/GaAs HBTSC prototype

With an upright cell growth, the base is accessed from the top and its metallic contact must be deposited in a groove parallel to the emitter contact. This increases the shadow factor of both junctions, and a trade-off must be found between shadow factor and series resistance. In the resulting cell, the emitter has been etched away in some regions to access the base, but wet chemical etching of GaInP is difficult to control, not very homogeneous and challenges with severe underetching, which leads to large margins around the contacts, reducing the top cell area and making it impossible to use an interdigitated contact design.

In this work, we use an inverted cell growth to mitigate these problems. The base contact is reached from the bottom of the structure and therefore the metallic fingers of the base can be deposited exactly beneath the front metal fingers, further reducing the shadow factor. We also include here another improvement with respect to the upright structures reported in [2], [3]: the base material is $Al_{0.1}Ga_{0.405}In_{0.495}P$ instead of $Ga_{0.505}In_{0.495}P$, increasing the base bandgap for better decoupling of top and bottom junctions within the transistor structure [6].

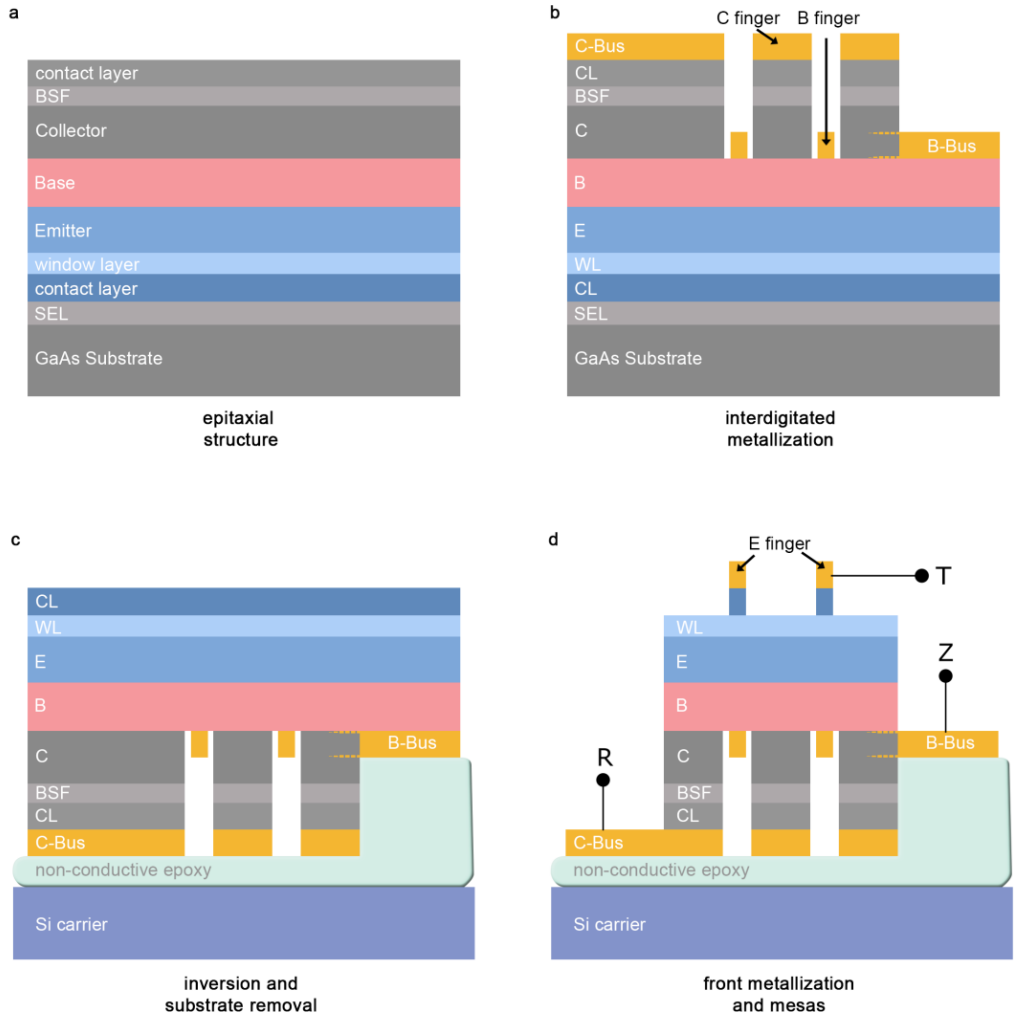

Fig. 2. Layer structure (a) and processing steps (b-d) on the inverted HBTSC prototype.

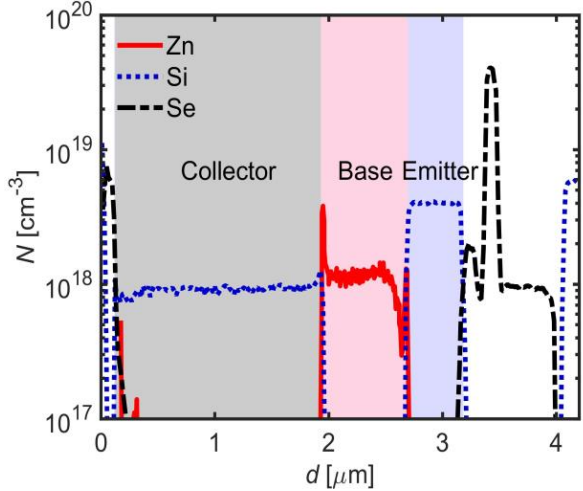

Fig. 3. Dopant atom concentration ($N$) as a function of depth from the surface of the epitaxy ($d$) (in structure as grown). SIMS measurement was carried out by Loughborogh Surface Analysis Ltd. UK.

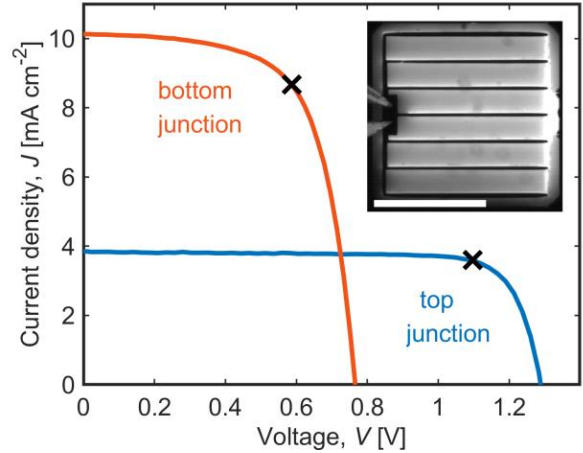

Fig. 4. *J-V* curves of top and bottom junction under illumination (AM1.5G) while the other junction is biased at the global maximum power point (crosses). Inset: electroluminescence image of the top junction at a current density equivalent to 20 suns. Scalebar is 1 mm.

## II. METHODS

Fig. 2 shows the semiconductor layers (a) and principal processing steps (b-d) of the cell presented in this work. The inverted semiconductor layer structure represented in fig. 2a was grown by metal-organic vapor-phase epitaxy (MOVPE). The emitter is nominally 550 nm thick, the base 800 nm and the collector 1800 nm.

In step b, the contact layer, back surface field (BSF) layer and collector are etched to form a groove for the base contact. To form the base contact, Au/Zn/Pd/Au are evaporated inside this groove. AuGe/Ni/Au are evaporated on the remaining area, which serves as the collector contact and as back reflector for the thin-film structure. This way, the base and collector contacts are interdigitated (the base busbar is actually perpendicular to the base fingers; in the scheme it appears parallel for simplicity). These two contacts are annealed at 420°C for 3 min. Then, in step c, the device is inverted and glued with an epoxy to a semi-insulating silicon substrate and the GaAs substrate is etched, revealing the epitaxial stack from the substrate side. AuGe/Ni/Au are evaporated as front contact and annealed at 180°C for 1 hour to avoid decomposition of the epoxy [8]. Then, mesas are etched, revealing the base and collector contacting busbars for wire-bonding (step d). The cell has no anti-reflection coating.

## III. EXPERIMENTAL RESULTS

Fig. 3 shows the doping profile of the epitaxial structure from a Secondary Ion Mass Spectroscopy (SIMS) measurement. Collector and base are doped with $10^{18}$ atoms cm$^{-3}$ Si and Zn, respectively, and the emitter is doped with $4 \cdot 10^{18}$ Si atoms cm$^{-3}$. There is a spike of Zn concentration at the interface between base and collector and a depletion between base and emitter, indicating a diffusion of Zn atoms towards the collector during growth, which accumulate at the interface between base and collector and do not propagate significantly into the latter. The resulting average Zn concentration in the base is lower than optimal [6].

Fig. 4 shows the current density ($J$) - voltage ($V$) curves of our first inverted HBTSC prototype when illuminated under an AM1.5G spectrum. The solar simulator was calibrated using an external reference cell made of the same materials. Here, we show the curves of top and bottom junction taken while the other junction was biased at its maximum power point. This is to give an outline of the device performance before showing the more detailed, three-dimensional *J-V-V* and Power (*P*)-*V-V* plots. The top junction reaches an open circuit voltage ($V_{OC}$) of 1.28 V and a short circuit current ($J_{SC}$) of 3.86 mA cm$^{-2}$. Its fill factor (*FF*) is 80%. The bottom junction has a $V_{OC}$ of 0.76 V, a $J_{SC}$ of 10.1 mA cm$^{-2}$ and a *FF* of 66%. An electroluminescence image of the top junction being biased to achieve a current density equivalent to 20 suns is being shown as an inset in fig. 4. In this photograph, the emitter bus is on the left, the base bus is on the right and the fingers of the base are aligned to the fingers of the emitter. The areas above the base contacts can be identified by a strong brightness, whereas the emitter fingers and bus appear dark. The fact that the brightness is significantly higher where the contacts are, suggests that the cell suffers from a high resistivity in the base layer (crowding effect) at the current level corresponding to 20 suns.

Fig. 5 shows the dark *J-V* curves of the junctions, together with their fit to the two-diode model. The top junction has $m_2 = 2.25$ with a dark recombination current density $J_{02} = 7 \cdot 10^{-13}$ A cm$^{-2}$. The bottom junction has $m_2 = 3.5$ and $J_{02} = 1.4 \cdot 10^{-6}$ A cm$^{-2}$. These ideality factors suggest that in both cases the recombination in the space charge region is dominating, while the bottom junction's ideality factor is far from that of an ideal junction. The series resistances ($r_s$) extracted from the dark curves are lower than 1 Ω cm$^2$.

Fig. 6 shows the external quantum efficiency (EQE) of the two junctions. The cell has no anti-reflection coating. While one junction was measured, the other junction was in open circuit. Note that the EQE overall is low compared to the theoretical absorbance of the layers, especially in the top junction. This was not the case for previous devices where the base was made of GaInP [2], which points to the AlGaInP alloy as cause for the poor carrier collection. Besides, the bottom junction has a non-zero EQE in the wavelength range of the top junction. This indicates luminescent coupling between top and bottom junctions, which is an indication of good material quality in the GaInP emitter.

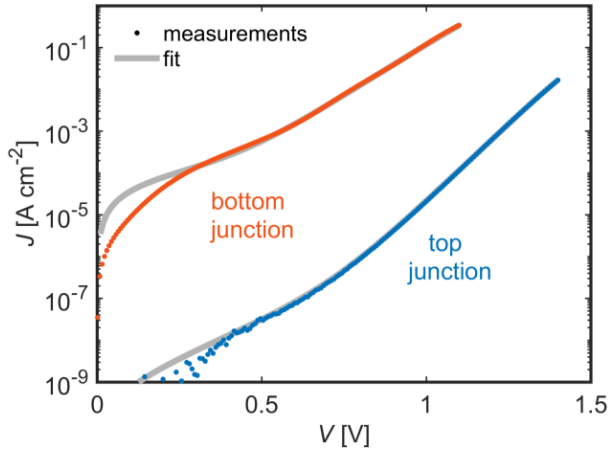

Fig. 5. *J-V* curves of top and bottom junction in the dark, together with their two-diode model fit.

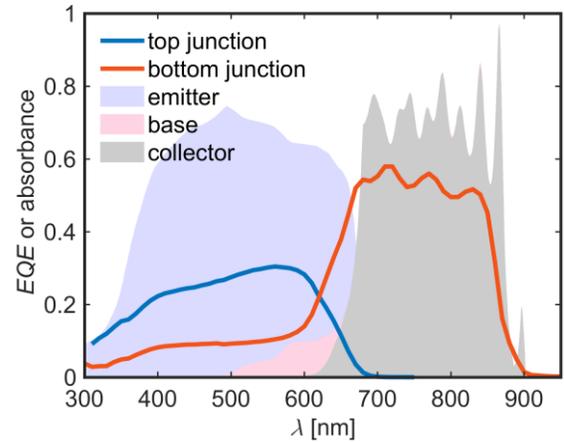

Fig. 6. External quantum efficiency (EQE) as a function of wavelength ($\lambda$) (lines), together with the theoretical absorbance of each layer (colored areas). The cell has no anti-reflection coating.

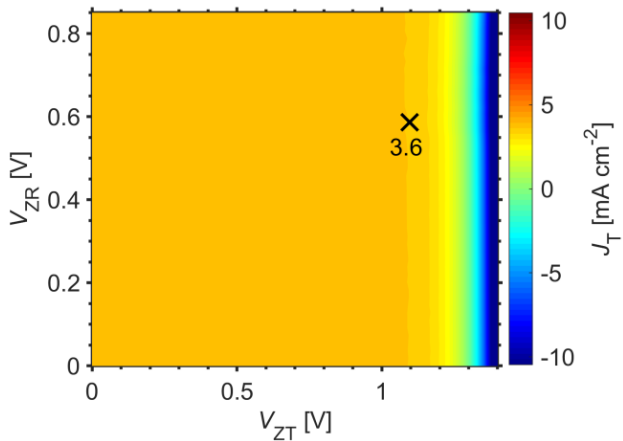

Fig. 7. Top junction current density ($J_T$) as a function of top junction voltage ($V_{ZT}$) and bottom junction voltage ($V_{ZR}$). The cross marks the maximum power point. The cell was illuminated (AM1.5G) and kept at 25 °C.

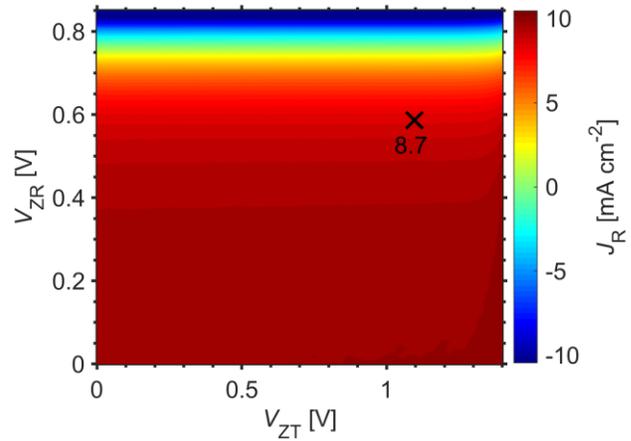

Fig. 8. Bottom junction current density ($J_R$) as a function of $V_{ZT}$ and $V_{ZR}$. The cross marks the maximim power point. The cell was illuminated (AM1.5G) and kept at 25 °C

Figs. 7 and 8 show the three-terminal (and therefore three-dimensional) *J-V* curves of the cell from a measurement carried out with two source-meters: the cell was illuminated (AM1.5G) and kept at 25 °C. The bottom junction was biased with a fixed voltage ($V_{ZR}$) while the top junction voltage ($V_{ZT}$) was swept and top junction current and bottom junction current were measured simultaneously. Fig. 7 shows the resulting top junction current density ($J_T$) and fig. 8 the bottom junction current density ($J_R$), both as functions of $V_{ZR}$ and $V_{ZT}$. Although they share a thin base layer, both $J_T$ and $J_R$ are independent from the bias of the other junction, when biased in the range from short circuit to their open-circuit voltage.

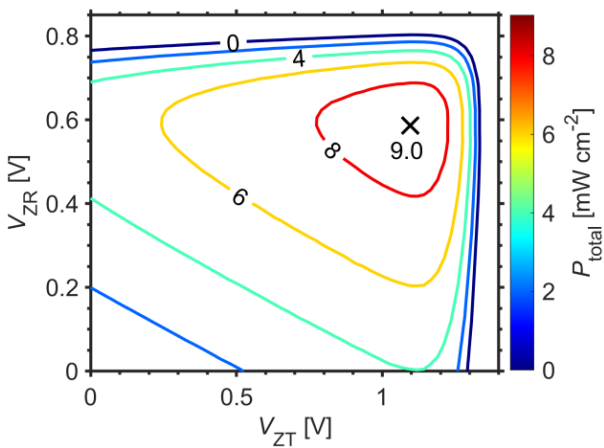

Fig. 9. Total power output as a function of $V_{ZT}$ and $V_{ZR}$. The cell was illuminated (AM1.5G) and kept at 25 °C. The cross marks the maximum power point.

The power density of the two junctions can be added up to calculate the total power density ($P_{total}$) for every set of voltages ($V_{ZT}$, $V_{ZR}$). Fig. 9 shows $P_{total}$ as a function of $V_{ZT}$ and $V_{ZR}$ of the illuminated cell. The maximum power point of 9.0 mW cm$^{-2}$ is reached at $V_{ZT}$ = 1.10 V and $V_{ZR}$ = 0.59 V.

## IV. Discussion

With the inversion of the HBTSC, we have achieved a technological breakthrough. This is mainly because in the inverted design, the base contact is achieved by etching through the GaAs collector from the bottom and not through the GaInP from above. This way, the fingers of emitter and base contacts could be stacked one on top of the other. The shadow factors have decreased from 0.47 (top) and 0.19 (bottom) in the upright structure [2] to 0.14 (top) and 0.19 (bottom) in the inverted structure, while keeping the series resistance at a similar level. As contacts are interdigitated in the inverted design and therefore the distance that charge carries have to travel from their point of generation to the metal contacts does not depend on the device area, the inverted cell design can be scaled to large areas while keeping the series resistance low. In the upright design, low series resistance could only be achieved for small device areas. In this case the cell has been attached to a plain Si carrier, but in the future, it will be possible to stack it onto other cells to form more complex multi-junction solar cells. In particular, for the case of a tandem with a silicon cell, we calculated a higher limiting efficiency than the series-connected version of the same materials [4].

The performance of this first prototype has room for improvement. The $FF$, $I_{SC}$ and $V_{OC}$ values are lower than in our previous upright prototypes which achieve 28% AM1.5G efficiency [2], [3]. We find that this can be attributed mostly to the choice of AlGaInP as base material instead of GaInP, rather than to the intricated processing of the inverted device. The main problems in the device performance are the low EQE, especially noticeable in the top cell, which results in a low short-circuit current, and the high recombination current and low $V_{OC}$ of the bottom junction. The low EQE is most likely because of the high emitter doping and the irregularities in the AlGaInP:Zn doping profile. High doping of the AlGaInP quaternary material is known to be challenging and the out-diffusion of Zn in this material has already been pointed out in the literature [9]. In the case of our HBTSC prototypes, for the same nominal doping a GaInP base has a measured dopant concentration of 8x10$^{18}$ cm$^{-3}$ [2], whereas this AlGaInP base has 1x10$^{18}$ cm$^{-3}$ and even 2x10$^{17}$ cm$^{-3}$ in the depleted region close to the emitter (see fig. 4).

With respect to the high recombination in the bottom junction, it is interesting to note that it is accompanied by a high ideality factor of 3.5. This could again be related to the doping problems of the AlGaInP base layer, in this case most likely to the accumulation of Zn atoms at the base/collector interface that results in a poor-quality junction. Since the introduction of an AlGaInP base seems to have created some problems that were not found in previous HBTSC prototypes with a GaInP base, it seems advisable to use GaInP in the future in spite of it being less optimal from the point of view of the HBTSC theoretical model [6], unless a breakthrough in MOVPE growth of the AlGaInP:Zn material is achieved.

## V. Conclusions

We have implemented an inverted design for the HBTSC and transferred it onto a Si carrier, paving the way towards substrate reuse and III-V–on-Si mechanical stacking of this technology. With the inverted design, the shadow factor is reduced, making the cell scalable to large-area devices. We present EQE measurements and $J$-$V$ curves of prototypes, showing that the inverted structure works as a double-junction solar cell. To improve the performance of the device it will be necessary to address the challenge of achieving a high doping in the AlGaInP base layer or substitute it by a GaInP layer.